# A Low-Voltage Retarding-Field Mott Polarimeter for Photocathode Characterization

J. L. McCarter,[a,b] M.L. Stutzman,[a,*] K.W. Trantham,[c,1] T.G. Anderson,[e] A.M. Cook,[d,2] and T.J. Gay[e]

[a] *Thomas Jefferson National Accelerator Facility, 12050 Jefferson Ave. Suite 500, Newport News, VA 23606, USA*
[b] *Department of Physics, University of Virginia, Charlottesville, VA 22901, USA*
[c] *Department of Physics, Fort Hays State University, Hays, KS 67601, USA*
[d] *Monmouth College, Monmouth, IL 61462, USA, and the US DOE SULI program,*
[e] *Behlen Laboratory of Physics, University of Nebraska Lincoln, Lincoln, NE, 68588-0111, USA*

* corresponding author
[1] present address: Bruner Hall of Science, University of Nebraska-Kearney, 2401 11th Ave., Kearney, NE 68849, USA
[2] present address: Fermi National Accelerator Laboratory, Batavia, IL 60510, USA

## Abstract

Nuclear physics experiments at Thomas Jefferson National Accelerator Facility's CEBAF rely on high polarization electron beams. We describe a recently commissioned system for prequalifying and studying photocathodes for CEBAF with a load-locked, low-voltage polarized electron source coupled to a compact retarding-field Mott polarimeter. The polarimeter uses simplified electrode structures and operates from 5 to 30 kV. The effective Sherman function for this device has been calibrated by comparison with the CEBAF 5 MeV Mott polarimeter. For elastic scattering from a thick gold target at 20 keV, the effective Sherman function is 0.201(5). Its maximum efficiency at 20 keV, defined as the detected count rate divided by the incident particle current, is $5.4(2) \times 10^{-4}$, yielding a figure-of-merit, or analyzing power squared times efficiency, of $1.0(1) \times 10^{-5}$. The operating parameters of this new polarimeter design are compared to previously published data for other compact Mott polarimeters of the retarding-field type.

## Introduction

Nuclear physics experiments have come to rely on high polarization electron beams, and place stringent demands on the electron source. In particular, beam requirements at Thomas Jefferson National Accelerator Facility's CEBAF (Continuous Electron Beam Accelerator Facility) include polarization over 80% and average current capability of at least 100 µA. Access to the CEBAF polarimeters for photocathode research is constrained by the experimental schedule. Previous offline photocathode polarization research at Jefferson Lab was performed using a 100 kV vertical electron gun and conventional Mott polarimeter, which required extensive radiation shielding and a personal safety system. We describe here a simple, load-locked, low-voltage polarized electron source used in conjunction with a newly designed compact, retarding-field Mott polarimeter which enables photocathode pre-qualification for the CEBAF injector as well as polarization characterization of novel photocathode materials.

# Polarized Electron Source

The polarized electron source is an ultra-high vacuum system where GaAs photocathodes are heated then activated to a negative electron affinity state using cesium and an oxidant. When illuminated with circularly polarized light at energies just over the band gap, activated GaAs emits longitudinally polarized electrons.[1,2] Figure 1 shows a SIMION[3] model of the beam trajectory through the source with typical bias voltages for each electrostatic element noted. The photocathode (element a in Figure 1) is biased at -268 volts with respect to the Mott polarimeter target using a battery bias box, and photocurrent is monitored with a picoammeter. Since Mott scattering detects an asymmetry for transversely polarized electrons, the spin direction of the initially longitudinally polarized electron beam direction must be bent 90°. This is accomplished using an electrostatic bend (elements b and c) of the design developed by Al-Khateeb et al.[4]. The beam is then focused and steered using one split lens (elements f and g) and two cylindrical lenses[5] (elements e and h) to the Mott polarimeter for polarization analysis.

For photoemission, we use a fixed (773 or 840 nm) or variable (~770-780 nm) wavelength laser, or a monochromator which produces un-collimated light at wavelengths from 650 to 850 nm. An x-y translational stage allows movement of the laser beam while maintaining normal incidence to the photocathode, and an optical attenuator system varies laser power and subsequent photocurrent. The final optical element is a quarter-wave plate to circularly polarize the light. The computerized data acquisition program controls a laser shutter and an insertable half-wave plate which is used to change the helicity of the circularly polarized laser light and to help cancel instrumental asymmetries.

New photocathodes are introduced into the source using a load-lock system, and a bake of the load-locked bellows at 250°C for 12 hours allows samples to be changed within a day. The photocathode is mounted on a hollow stainless steel "stalk" where the GaAs is heated to a surface temperature of ~550°C for two hours using an external heater prior to chemical activation with Cs and the oxidant $NF_3$ (though $O_2$ could be used instead). An 8" diameter stainless steel chamber houses the polarized source and is pumped with a combination of ion and non-evaporable getter pumps[6]. To achieve pressure in the low UHV regime, the polarimeter was initially baked in a hot air oven for 30 hours at 200°C, which is the limit for the Teflon insulators in the lens transport system and the CEM support structure. During the bake, a sheet metal wall separates the hot air oven into two sections so that the source chamber can be heated to nearly 250°C to further reduce the water vapor pressure and activate the NEG pumps to approximately 60% of their maximum pump speed. The combination of NEG and ion pumps typically achieves pressure in the low $10^{-11}$ Torr range, leading to a very long photocathode lifetime, with reactivations necessary every few months.

The polarized source is coupled to the Mott polarimeter with a 3.2 mm diameter aperture on centerline to define the beam. The source lens system is separated from the Mott lens transport system by 15 cm, and an isolated planar electrode can be inserted after the aperture before the Mott transport system to monitor the current that enters the lens transport system.

# Mott Polarimetry

Our Mott polarimeter has a particularly simple design, shown in Figure 2, with no electrode structures except the inner high-voltage hemisphere and the grounded outer hemisphere, which in turn supports simplified retarding-field grids.  Its hemispherical structure is similar to a "mini-Mott" design reported earlier,[7] but is smaller and simpler, eliminating guard rings and other ancillary electrodes.  It is also similar to a micro-Mott design discussed briefly by Ciccacci et al.[8] that is somewhat larger.  The electrode structure supports voltages at least as high as our 30kV power supply maximum.

Many parameters have historically been used to characterize Mott polarimeters.  Because incident electron currents are often low, the polarimeter's detection efficiency, defined as the electron detection rate divided by the incident electron current, $I/I_o$, is important.  In addition, experiments requiring spin analysis of scattered electrons often place severe spatial constraints on the size of the polarimeters that can be used.[9,10]  So-called "micro-Mott polarimeters," developed largely at Rice University by Dunning and co-workers over the last two decades,[11,12,13,14] solve these two issues simultaneously by reducing polarimeter size; as a rule of thumb, the detection efficiencies of Mott polarimeters vary inversely with their size, since the chief factor in determining efficiency is the effective solid angle subtended by the electron detectors at the Mott scattering target.   The figure of merit, η, for a Mott polarimeter is inversely proportional to the square of the time required to measure polarization to a given statistical accuracy,[9,15] and is defined as

$$\eta \equiv (S_{eff})^2 \times I/I_o \, , \tag{1}$$

where $S_{eff}$ is the effective Sherman function or polarimetric analyzing power given by

$$S_{eff} = A/P_e \, , \tag{2}$$

$P_e$ is the electron polarization, and

$$A = \frac{R-L}{R+L} \tag{3}$$

is the scattering asymmetry, with the values of R and L corresponding to the count rates in the "left" and "right" electron detectors of the Mott polarimeter.  Generally speaking, as the detection solid angle, and thus the ratio $I/I_o$, increases, $S_{eff}$ decreases.

## Polarimeter Design

Electrons that enter the polarimeter are accelerated to energies from 5 to 30 keV between two hemispherical stainless steel electrodes supported on a ceramic insulator (see Figure 2).  Electrons scatter from a gold target (5 microns of gold plated on a copper cylinder) inside the inner hemisphere.  In principle, the target could be biased negatively relative to the inner hemisphere to suppress noise due to ions accelerated into the detectors.[16]  This was not done, as no ion-related noise was observed.  The vacuum chamber serves as adequate radiation shielding at 30 kV for typical operating currents up to 100 nA on target.  Scattered electrons are decelerated in the gap between the inner and outer hemispheres and detected with channel electron multipliers (CEMs)[17] , each subtending 0.27 sr, centered at 120°. To reduce the chance of electrical discharge, the outer surface of the inner hemisphere was highly polished (mirror finish with 5 micro-inch rms surface roughness), and the aperture holes in both hemispheres were rounded and polished.  Two gold mesh[18] grids in front of each CEM, separated by 3.5 mm, establish a spatially well-defined retarding potential volume and reject inelastically-scattered electrons.  The grids are affixed to aluminum support rings using Aerodag[20] and isolated by ruby balls.   As the retarding potential is increased negatively from ground, electrons that have lost energy through inelastic scattering are increasingly excluded from the measurement;  when the retarding potential energy approaches that of the incident beam kinetic energy, only the elastically scattered electrons are detected.  In this paper, we will use ΔE to refer to the greatest energy a scattered electron can lose and still be detected.  Thus, for an incident beam with kinetic energy K and a retarding voltage on the grids equal to V, $\Delta E = K - |e|V$ .  The two-grid retarder design has been found to provide better discrimination against inelastically scattered electrons at small values of ΔE.[19]  The polarimeter has four detectors: the right/left pair is aligned to measure the Mott scattering asymmetry and the up/down pair can be used to measure any out-of-plane polarization due to physical mechanisms, instrumental asymmetries, or polarimeter misalignment.  Electrostatic tube lenses and deflectors[20] steer and focus the incident electron beam into the polarimeter entrance aperture.

## Signal Processing

In order to determine the efficiency of the Mott analyzer accurately, it is important to ensure that the signal pulses are associated with true target-scattered electron events, and that electronic dead time does not affect the result.  Dead time issues were addressed by operating in the regime where count rates increased linearly with the incident beam current, and where efficiency was steady.  This occurred for target currents less than 50 pA at 5 kV target bias and count rates less than 1 MHz (see Figure 3). The operating voltages for the CEMs were determined by both finding the point where a 100V increase in bias produced less than a 10% increase in count rate and using an oscilloscope to ensure that the primary pulse peak height did not change.  The CEM high voltage[21] bias boxes are outside the vacuum chamber, and each channel is in a separate metal housing to reduce cross-talk.  The capacitively coupled CEM output signal is amplified with a pre-amp[22] placed immediately adjacent to the bias box.

The discriminator[23] threshold was determined by measuring both the asymmetry and signal-to-noise ratio as a function of threshold voltage with $\Delta E \approx 150$ eV to eliminate the high count rate from scattered electrons with the largest energy losses.  Figure 4 shows that discriminator thresholds of at least 250 mV are needed for the signal/noise ratio and asymmetry to be independent of discriminator threshold; thresholds of 400 mV were typically used during data acquisition.  Peak pulse heights were typically over 1 V after amplification.  The TTL pulses from the discriminators were counted via a computerized DAQ program.

## Asymmetry Measurement

Polarized electrons are emitted from the GaAs photocathode in two opposite polarization states depending on the handedness of the incident circularly-polarized light.  Measuring count rates in both the left (L) and right (R) detectors during both polarization states, designated by subscripts of 1 and 2, allows cancelation of many of the instrumental asymmetries.[10] With the definition $N^+ = \sqrt{L_1 R_2}$ and $N^- = \sqrt{L_2 R_1}$, the asymmetry is given by

$$A = \frac{N^+ - N^-}{N^+ + N^-}. \tag{4}$$

## Results

### Efficiency

Efficiency, $I/I_0$, was measured by first biasing the target at +300V and measuring the incident current with a picoammeter, then biasing the target at operating voltages up to 30 kV and measuring CEM count rates. Maximum efficiency, with essentially no rejection of inelastically-scattered electrons, is shown as a function of target bias in Figure 5a. The monotonic decrease of efficiency with increasing target voltage is a result of lowered electron scattering cross sections at higher incident energies. The efficiency was measured using the same polarized electron beam as is used for the asymmetry measurement, and was determined as a function of ΔE for various target biases as shown in Figure 6a.

To verify that the target current at 300V accurately represents target current at higher biases, current was measured as a function of target voltage using batteries up to 300V and using a high voltage power supply[24] with nanoamp current sensitivity up to 7kV (see Figure 7). The slight increase in target current with target bias can be attributed to an increase in the number of secondary electrons produced at the target and upstream apertures that return to the target at higher bias.

### Effective Sherman Function

The effective Sherman function, $S_{eff}$, was determined by generating electron beams from the same photocathode material and laser wavelength as used in Jefferson Lab's CEBAF polarized electron source,[25,26] and dividing the measured asymmetry by the known beam polarization. These strained superlattice GaAs photocathodes,[27] which consist of 14 pairs of layers of GaAs (4 nm) on GaAsP (3 nm), generate electron beams with polarization of 84% (±1% statistical ±1% systematic) when illuminated with 778 nm light, as measured by the CEBAF 5 MeV Mott polarimeter[28] and corroborated by the four polarimeters in Jefferson Lab's three experimental halls. This value is reproducible across the photocathode diameter and between wafers. The $S_{eff}$ vs. target bias is shown in Figure 5b. A linear weighted average fit of $S_{eff}$ versus ΔE, using the range ΔE = 0 to 115 eV, was used to determine $S_{eff}$ for ΔE=0 since the count rates when ΔE=0 are quite low. Figure 6b shows both data and fit for 20 and 30 kV target bias. Backgrounds subtractions were made for both the residual rate with the light off and the residual rate when the retarding field exceeded that required to exclude elastically-scattered electrons from the detectors. The error bars reflect statistical uncertainty in the asymmetry measurement as well as the ±1% systematic and ±1% statistical uncertainty (added linearly) in the CEBAF polarization measurement. The average Sherman function for 20 kV and ΔE=0 was found to be 0.201 ± 0.004. The polarization in the vertical plane was measured for the same superlattice photocathode material using the vertical CEMs. At 20 kV target bias, the vertical component of asymmetry was 6.9% that of the horizontal component, corresponding to a polarimeter misalignment of 4°. Including this vertical component would increase $S_{eff}$ by only 0.0005, and was not included in the remaining calculations for the paper.

### Photocathode material comparison

As a verification of the determination of $S_{eff}$, polarization measurements were also made using "bulk" GaAs wafers diced from a single crystal,[29] and "strained layer" photocathodes with a single 100 nm thick GaAs layer grown on a lattice-mismatched substrate.[30] Figure 8 shows polarization measurements vs. target bias for the two materials, with data shown for several cycles of photocathode replacement and beam re-steering. The variation between nominally identical samples gives an estimate of the random systematic error of the measurements, approximately ± 3% of the value. Polarization of electrons from strained GaAs measured at CEBAF is typically around 77%, consistent with the measurements from this polarimeter. The measured polarization of bulk GaAs can vary widely depending on factors such as the substrate thickness and surface conditions, and the measured polarizations near 30% for all target biases are within expectations.

### Wavelength dependence

Figure 9 shows the wavelength dependence of both polarization and quantum efficiency (QE) for a high polarization strained superlattice GaAs/GaAsP photocathode[31] from 725 to 825 nm. The broad peak in maximum polarization from 780 to 795 nm is evident, and the results from this polarimeter are in good agreement with data taken previously with the JLab 100kV vertical test stand Mott polarimeter using a wavelength tunable Ti-Sapphire laser, shown by the solid line. For the longest wavelengths, statistical error bars dominate due to the very low QE.

### Figure of Merit

Figures 5c and 6c show η as a function of target bias and ΔE. Since $I/I_o$ increases several orders of magnitude with ΔE and the Sherman function decreases by less than a factor of two over the same range, the highest η at is found at ΔE=268V, corresponding to the incident beam energy. The measured η was lower than that of comparable polarimeters, as a result of the previously noted decrease in efficiency, which outweighs the small increase in $S_{eff}$ of this design.

## Discussion

Table 1 compares the operating characteristics of all of the "micro-" and "mini-" Mott polarimeters reported in the literature to date for which operating parameters are given. For the sake of comparison, all operating characteristics were determined for high-voltage operation at 20 kV, except where noted. Listed are the maximum values of $S_{eff}$, $I/I_o$ and η reported, the $S_{eff}$ values corresponding to the maximum reported efficiencies and figures-of-merit, and the values of ΔE corresponding to all these quantities. Also tabulated are estimates of the geometric scattered electron acceptance solid angle per detector (corresponding to straight-line trajectories from the target), and the approximate cylindrical volume of each device.

Some general statements can be made as a result of these comparisons. Mini-Mott polarimeters are characterized by sizes of the order of $10^4$ cm$^3$, while the smallest micro-Mott polarimeters take up volumes less than 1.5 x $10^3$ cm$^3$. The mini-Mott polarimeters have generally higher values of $S_{eff}$ for all values of ΔE due to their more restricted angular scattering acceptance.

Despite the fairly large acceptance of the present polarimeter (smaller only than those of references [9] and [38]), both I/I$_o$ and the corresponding η are quite small. This may be due in large part to the fact that we were limited in this experiment to ΔE ≤ 270 eV; SIMION analyses do not indicate a significant non-geometric rejection of scattered electron trajectories by our analyzer.

More direct comparisons can be made between our polarimeter and those reported in references [13] and [14], as they provide $S_{eff}$, I/I$_o$ and η results as a function of ΔE. While the sizes and detection solid angles of these devices are comparable to ours, both of the other polarimeters used Th targets, whereas our target was Au. To facilitate a direct comparison, we have used the results of Oro *et al*.[32] and McClelland *et al*.,[33] who studied the values of $S_{eff}$ for both targets at 20 keV as a function of ΔE. Our scaled values of $S_{eff}$ are given in the last row of Table 1. Using the results of Browning *et al*.[34] and Czyżewski *et al*.,[35] we take I/I$_o$ to scale roughly as the atomic number of the target; our I/I$_o$ and η values extrapolated to Th are also shown in the last row of Table 1. With these assumptions in place, and for ΔE= 268 eV, our device has $S_{eff}$ = 17%, as opposed to values of 23% and 25% for refs. [13] and [14] (extrapolated to ΔE= 268 eV). Given that our detection solid angle is nominally 20% greater than those of the Rice polarimeters, this is not terribly surprising. However, the values of I/I$_o$ for the Rice detectors (again extrapolated to ΔE= 268 eV) are roughly 3 to 4 times larger than those for our device, with comparably larger values of η. This result is not understood at this time, but it does not present a serious problem in terms of studying high-current photocathodes.

## Conclusions

We have commissioned a simple micro-Mott polarimeter/polarized electron source system for photocathode characterization for which the chief benefits are rapid sample changes, simplicity of construction, versatility of operation, and small size. Its operation range is 5 to 30 kV, eliminating the radiation hazards present with Jefferson Lab's previous offline polarimeter. The polarimeter's analyzing power, or "effective Sherman function", $S_{eff}$, has been calibrated through a comparison with Jefferson Lab's CEBAF 5 MeV Mott polarimeter by measuring polarization from the same high-polarization photocathode material with both devices. The present design has analyzing power and efficiency comparable to early designs of micro-Mott polarimeters. In comparison with state-of-the-art designs, it has a comparable analyzing power, but significantly lower efficiency and subsequent figure-of-merit. This lower efficiency, which cannot be understood simply in terms of detector acceptance, is not a problem for this system, which is intended to characterize high-current photocathodes. The polarized source in conjunction with the compact, retarding-field Mott polarimeter is a valuable tool for off-line photocathode prequalification and novel photocathode polarization research for the Jefferson Lab Center for Injectors and Sources.

# Acknowledgements

Notice: Authored by Jefferson Science Associates, LLC under U.S. DOE Contract No. DE-AC05-06OR23177. The U.S. Government retains a non-exclusive, paid-up, irrevocable, world-wide license to publish or reproduce this manuscript for U.S. Government purposes. The Mott polarimeter was designed and built at the University of Nebraska under grants NSF PHY-0099363 and PHY-0354946.

# Figure Captions

Figure 1: SIMION model of beam transport through the polarized electron source showing lenses and typical voltages in cross section, with the three-dimensional inset showing detail of the photocathode and "pusher" electrostatic bend. The incident laser beam path and the SIMION modeled electron trajectory are labeled d and i respectively. The beam limiting aperture is labeled j, and k indicates the insertable planar electrode for current monitoring.

Figure 2: Scale cross section drawing of polarimeter showing: 1) 8"Conflat ® mounting flange with 2¾" ports for high-voltage bushings and feedthroughs; 2) insulating standoff and mounting plate; 3) outer hemisphere; 4) highly polished stainless steel inner hemisphere; 5) target screwed into high voltage electrode; 6) channel electron multiplier in housing attached to retarding-field grid assembly.

Figure 3: Efficiency ($I/I_0$) *vs*. target current at 5 keV target bias and $\Delta E=268$ eV (open circles). Efficiency error bars are dominated by uncertainty in the current measurement. Channel electron multiplier (CEM) count rate (closed circles) varies linearly for target currents below 100 pA and rates below 1 MHz while counting efficiency drops over 100pA or 1MHz (linear fit to data below 50 pA and extended as a guide to the eye). Count rate was kept below 1 MHz (vertical dotted line) during measurements to avoid electronic saturation effects.

Figure 4: Asymmetry (circles) and CEM count rate with beam on (dashed line) and beam off (dotted line) as a function of discriminator threshold. Ratio of beam-on to beam-off count rate is indicated by the solid line (see text). Data shown is for 5 kV with superlattice photocathode and 773 nm laser illumination.

Figure 5: Variation as a function of target bias of the a) efficiency, $I/I_o$ for $\Delta E=268$eV; b) effective Sherman function for $\Delta E=268$ eV (open circles) and extrapolated to $\Delta E=0$ eV (closed circles); and c) the figure of merit, η, for $\Delta E=268$eV.

Figure 6: Variation, as a function $\Delta E$, of a) efficiency, $I/I_o$; b) effective Sherman function, $S_{eff}$, with weighted linear fit for extrapolation to $\Delta E=0$ eV; and c) figure of merit, η. Filled circles correspond to target bias of 20 kV; open circles to 30 kV.

Figure 7: Target current as a function of target bias relative to that measured with the target grounded, with currents typically on the order of 100 nA (see text).

Figure 8: Measured electron polarization vs. target bias.  Solid line indicates superlattice polarization of 84% used to determine $S_{eff}$.  Strained layer data: squares measured February 2008, open diamonds July 2008, solid diamonds September 2008.  Bulk GaAs: open circles measured September 2008, solid circles June 2008.

Figure 9:  Polarization (closed circles) and quantum efficiency (open diamonds) plotted as a function of wavelength for a high polarization superlattice photocathode.  Polarization vs. wavelength data from 100 kV vertical test stand and Mott polarimeter are shown by the solid line.

**Table 1: Comparison of various mini-Mott and micro-Mott designs at 20 kV with Au targets.**

| Ref. | Laboratory | Max. $S_{eff}$ (%) | ΔE (eV) | Max. $I/I_o$ (10⁻⁴) | ΔE (eV) | $S_{eff}$ at max. $I/I_o$ (%) | Max. η (10⁻⁵) | ΔE (eV) | $S_{eff}$ at max.η (%) | ΔΩ (sr) | Volume (10³cm³) | Notes |
|---|---|---|---|---|---|---|---|---|---|---|---|---|
| 7  | Rice      | 26 | 0    | 14   | 1300 | 12 | ~2  | 1300 | 12 | 0.02 | 8.5 |   |
| 14 | Rice      | 23 | 400  | 53   | 1000 | 16 | 13  | 400  | 23 | 0.21 | 1.1 | a |
| 13 | Rice      | 21 | 300  | 94   | 1500 | 9  | 12  | 700  | 15 | 0.25 | 2.9 | b |
| 12 | Rice      | 11 | 1300 | ~20  | 1300 | 11 | 2.4 | 1300 | 11 | 0.11 | 2   |   |
| 11 | Rice      | 11 | 1300 | 22   | 1300 | 11 | 2.7 | 1300 | 11 | 0.09 | 4.2 |   |
| 36 | Münster   | 25 | 0    |      |      |    |     |      |    | 0.02 | 9.3 |   |
| 37 | Irvine    | 20 | 500  | 6.7  | 1000 | 14 | 1.4 | 1000 | 14 | ?    | ?   | c |
| 9  | Taiwan    | 13 | 700  |      |      |    | ~2  |      |    | 0.60 | ?   | d |
| 38 | Tokyo     | 13 | 600  | 195  | 1400 | 10 | 18  | 1200 | 10 | 0.57 | 1.2 |   |
| 39 | St.Pet.   |    |      |      |      |    | 4.5 |      |    | 0.06 | 1.3 | e |
| 8  | Edinburgh | 9  | 1300 |      |      |    |     |      |    | 0.06 | 2.2 |   |
| **This work** |  | 20 | 0 | 5.4 | 268 | 13.5 | 1.0 | 268 | 13.5 | 0.27 | 1.4 | |
| **This work Th adj.** |  | 27 | 0 | 6.2 | 268 | 17 | 1.8 | 268 | 17 | | | |

a. **Th target; 25 keV; max η occurs over range of ΔE from 400 to 1000 eV**
b. **Th target**
c. **U target**
d. **23 keV**
e. **30 keV; refs. (13) and (14) indicate little change in η between 20 and 25 keV at 1300 eV**

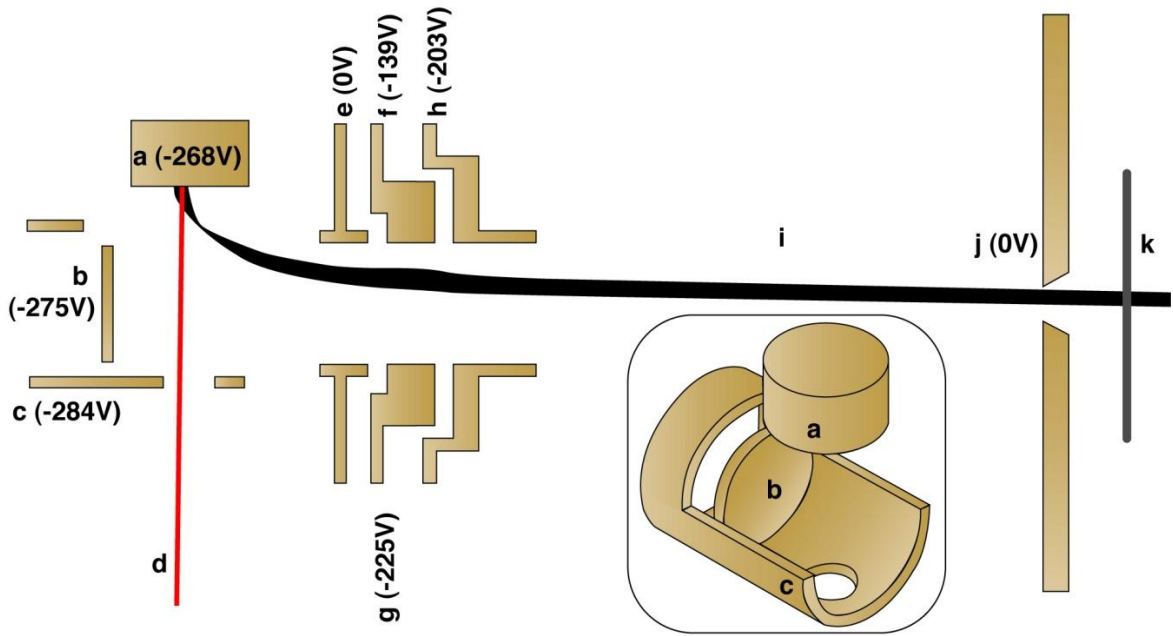

Figure 1

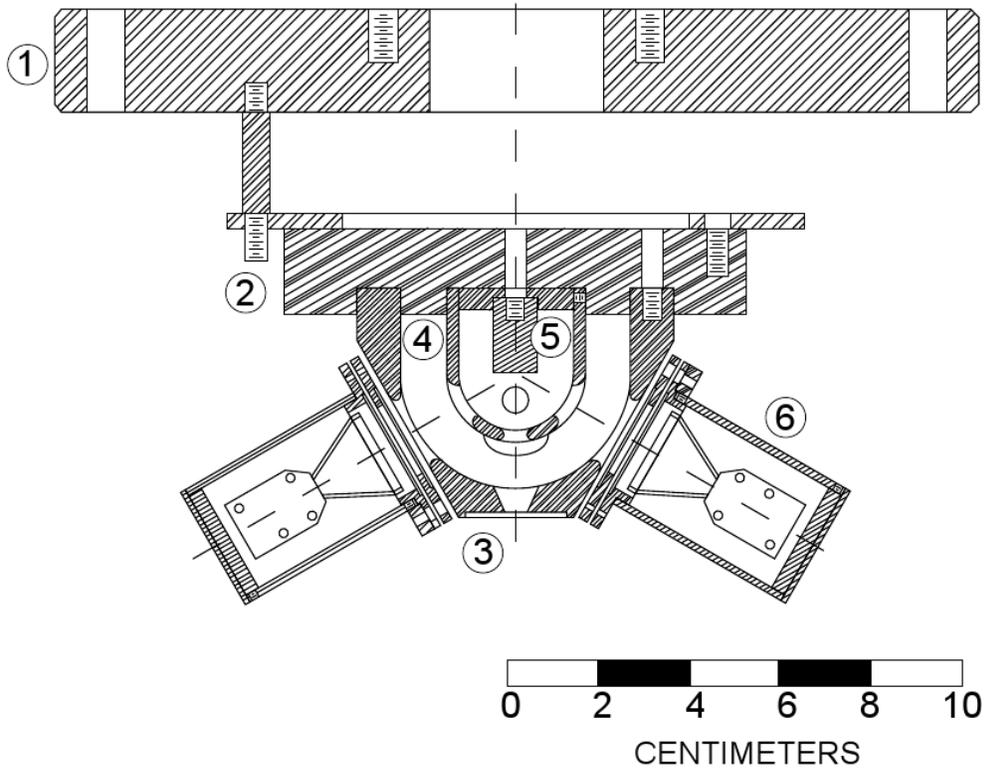

Figure 2

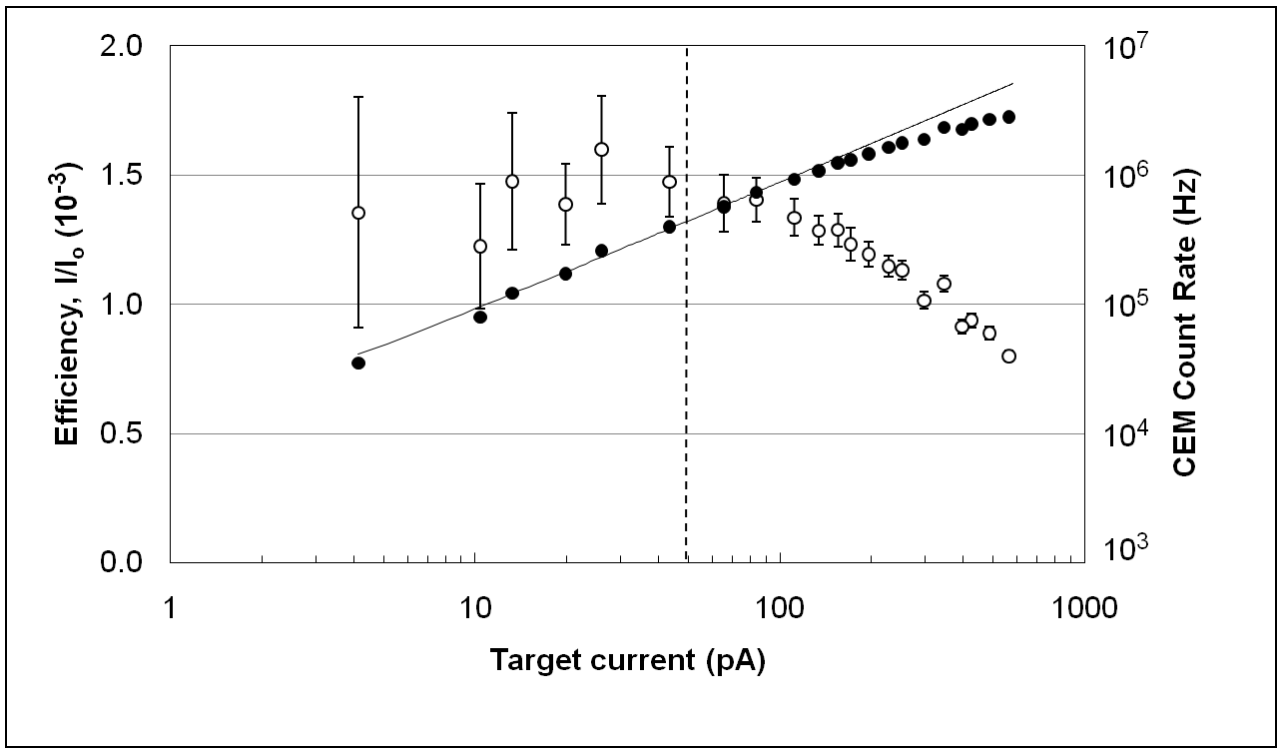

Figure 3

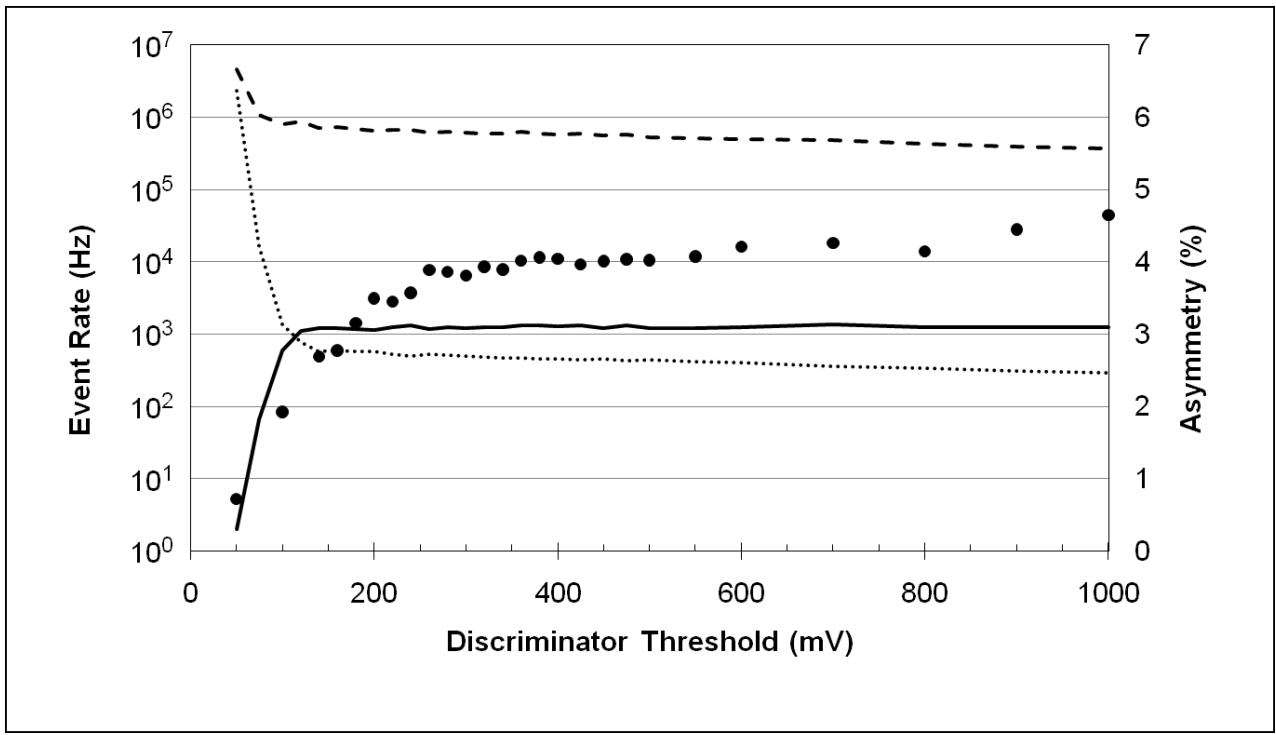

Figure 4

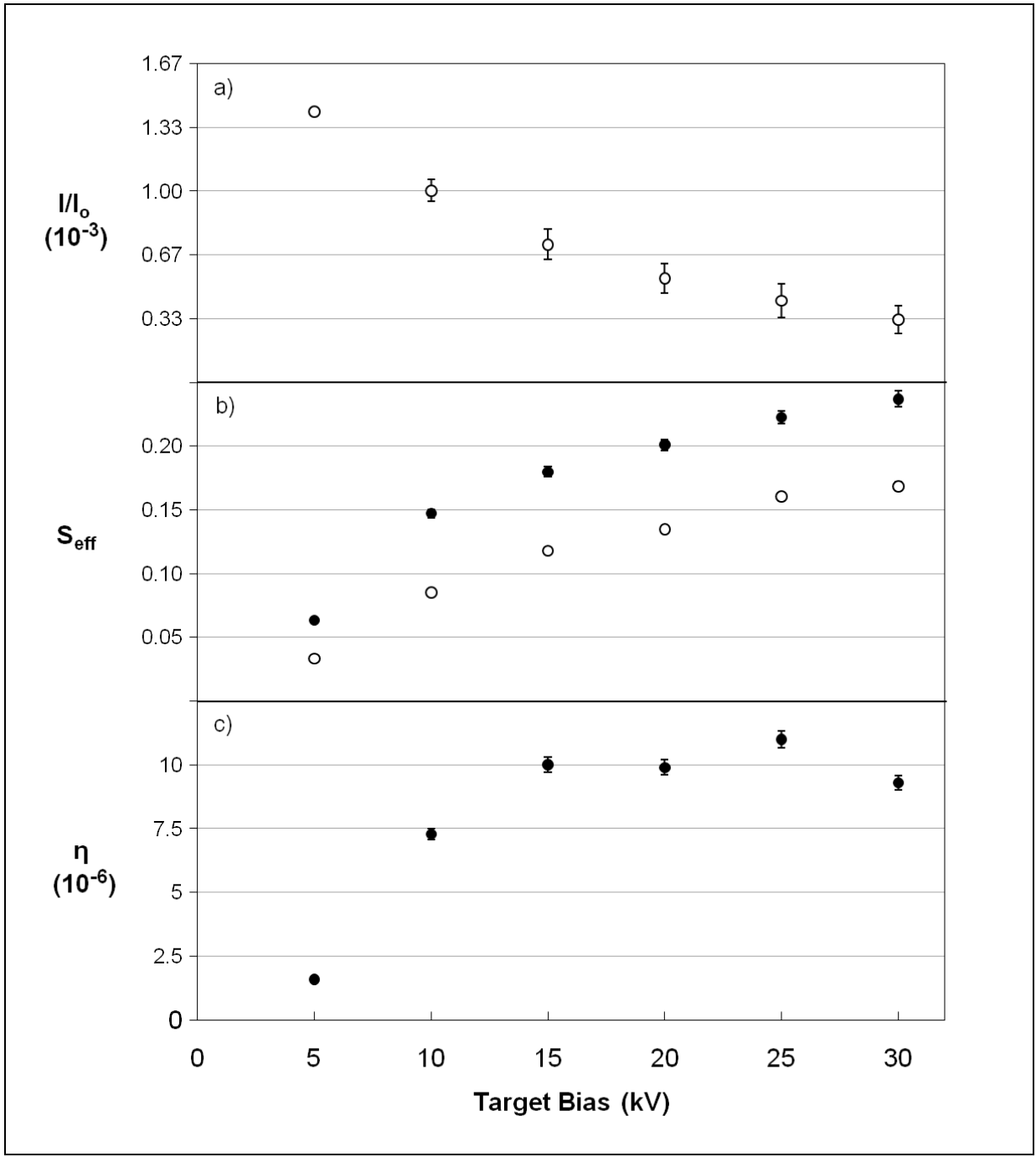

Figure 5

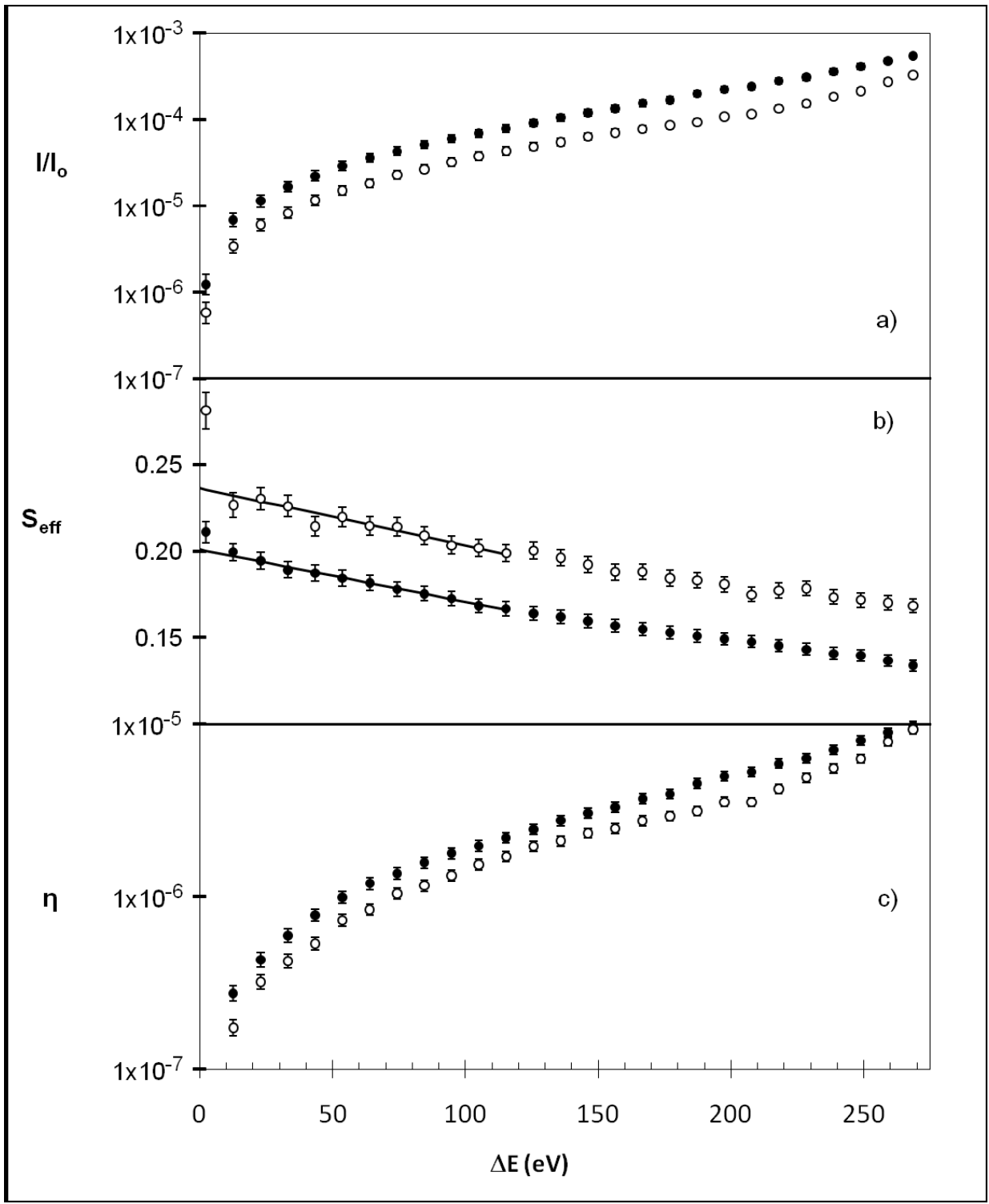

Figure 6

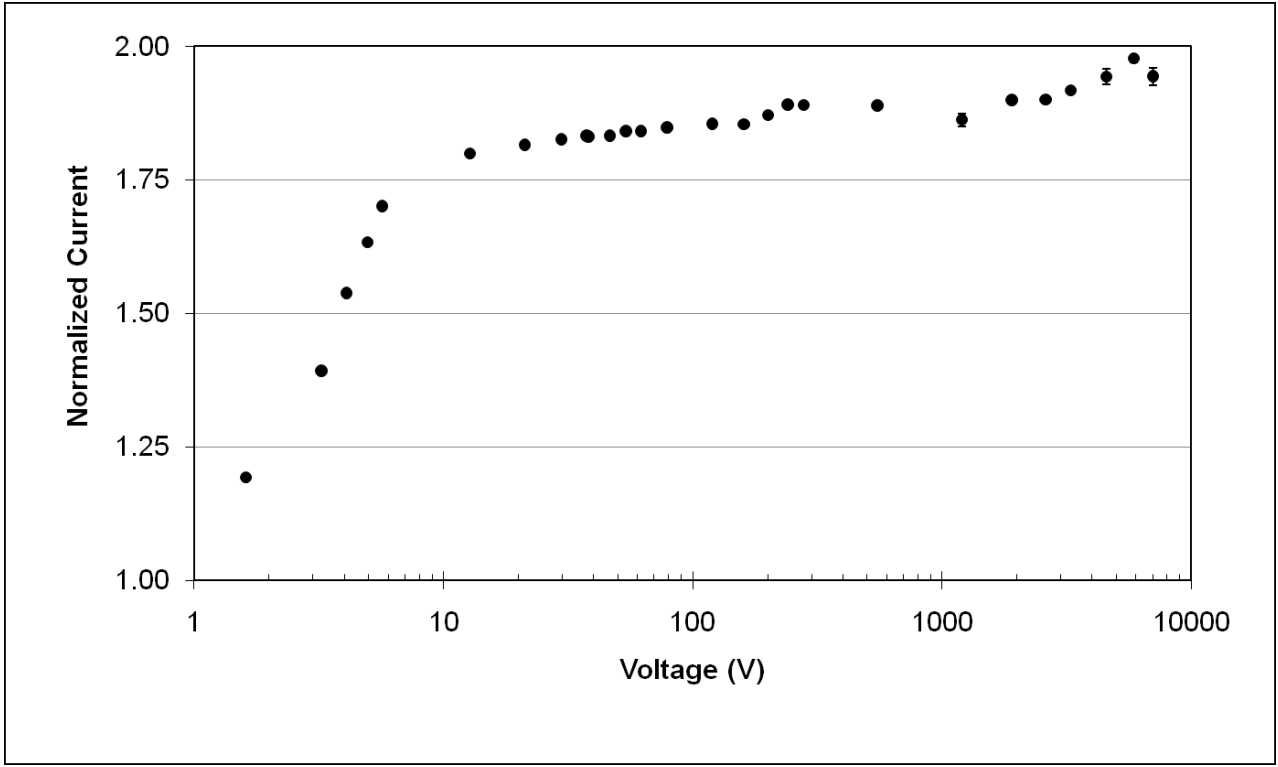

Figure 7

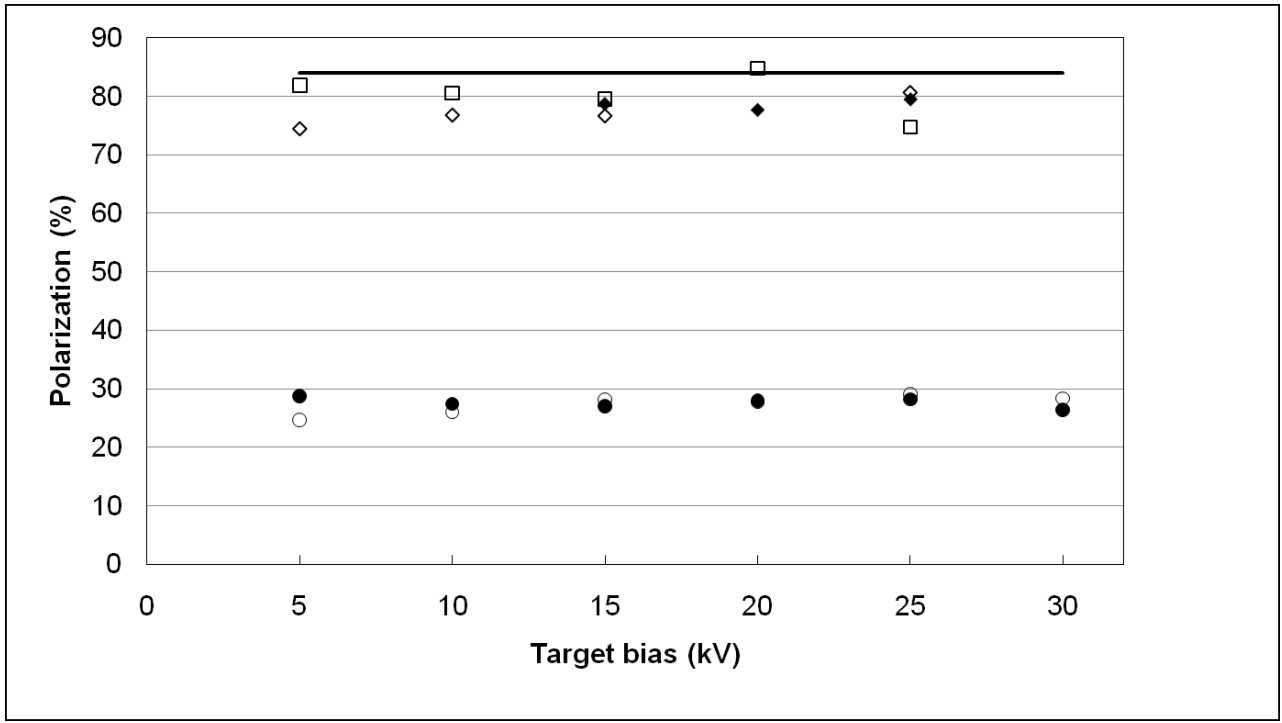

Figure 8

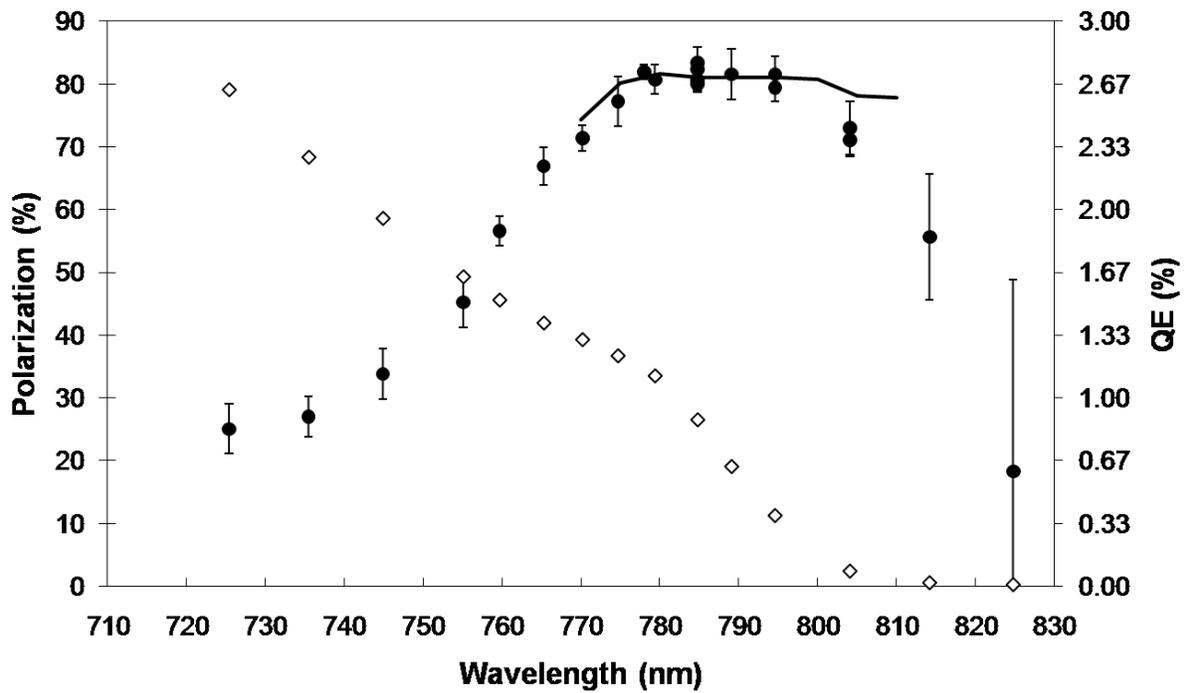

Figure 9